# Usability, Accessibility and Web Security Assessment of E-government Websites in Tanzania

Noe Elisa
Department of Computer Science
The University of Dodoma,
Dodoma, Tanzania

## ABSTRACT
In spite of the fact that e-government agency (ega) in Tanzania emphasize on the use of ICT within public institutions in Tanzania, accessibility, usability and web security vulnerabilities are still not considered by the majority of web developers. The main objective of this study is to assess the usability, accessibility and web security vulnerabilities of selected Tanzania e-government websites. Using several automatic diagnostic (evaluation) tools such as pingdom, google speed insight, wave, w3c checker and acunetix, this study assess the usability, accessibility and web security vulnerabilities of 79 selected e-government websites in Tanzania. The results reveal several issues on usability, accessibility and security of Tanzania e-government websites. There is high number of usability problems where 100% of websites were found to have broken links and 52 out of 79 websites have loading time of more than five (5) seconds for their main page. The accessibility results show that all 79 selected websites have accessibility errors and violate w3c Web Content Accessibility Guidelines (WCAG) 1.0. The results on web security vulnerabilities indicate that 40 out of 79 (50.6%) assessed websites have one or more high-severity vulnerability (SQL injection or cross site scripting-XSS) while 51 out of 79 (64.5%) have one or more medium-severity vulnerabilities (Cross site request forgery or Denial of Service). Based on these results, this study provides some recommendations for improving the usability, accessibility and web security vulnerabilities of public institutions in Tanzania.

## 1. INTRODUCTION
The use of information and communication technology (ICT) has become of great importance in delivering public services to the users. Over the past few decades, governments in the world have introduced electronic government as effective and efficient way of communicating services to their citizens, businesses, governments and employees. In the past few years, the government of Tanzania adopted the use of ICT in accelerating and enhancing communication to the citizens. In order to access online services provided by governments to their users, Webpages are the major interfaces. Despite the proliferation of e-government services, online services always face major challenges such as usability problem, accessibility issues and security breaches which may result into frustrating users from using it. For example, a study conducted to investigate the usability and user experiences of Kenya government websites identified that government websites are partially usable in the design layout perspectives whilst user experiences were poor and most users only revisit the sites as an obligation or lack of a better option [1]. Transforming government into e-government brings benefits such as cost-effective delivery of services, integration of services, reduction in administrative costs, a single integrated view of citizens across all government services and faster adaptation to meet residents' needs [2]. Websites and internet facilitate e-government in disseminating information and services to the users in today's information societies.

In Tanzania, e-government services were introduced in order to help citizens be able to access all government services electronically at the same time keeping transparency and ensuring interoperability. Tanzania E-government agency (ega) strategic plan (2012/2013-2015/2017) policy articulates the need for increasing the use of Information and Communication Technology (ICT) within public institutions, to enhance work efficiency and improve service delivery to the public [3]. When designing the websites for these public institutions, the basic design principles should be taken into consideration to make it easy for users to access services online. A website is considered to follow basic principles of web standards if it is effective, efficient, memorable, error free and provide satisfaction to its users [4]. Therefore the need for assessing e-government websites usability, accessibility and security in Tanzania is required in order to keep smooth operation of government electronic activities. This work has adopted the usability definition by Nielsen and Loranger [5] who defined the usability as an attribute of quality that refers to the promptness with which users learn to use something, the efficiency they attain while making use of it, how easy it is for them to remember how to use it, how error-prone it is and the level of satisfaction that they attain from using it. In [6] web accessibility is defined as a subset of usability and require that people with disabilities can use it[7]. Therefore, a website is said to be accessible if people with disabilities can perceive, understand, navigate, and interact with the Web, and be able to add their information to the web[8]. In this study, it is believed that if a web is accessible to the disabled people then it would benefit all users of the web. The disabilities considered in this study are those stated in [9][10] and might be blindness, low vision, color-blindness, inability to use a mouse, slow response time, learning disabilities, distractibility, inability to remember or focus on large amounts of information, Deafness, and hard-of-hearing.

Security in e-government websites is a mechanism for protecting adversaries from accessing sensitive information shared by its users. In [11] the authors tried to insist that the effective management of information security in e-government is a key factor as willingness and trust of the different users (citizens, businesses employees e.t.c) to use e-government services. While governments are trying to harness the use of ICT to simplify services delivery, little effort and attention have been devoted to ensure usability and accessibility standards are followed plus enforcing security of these services by minimizing the vulnerabilities.

The studies conducted to assess the factors affecting the adoption of e-government services found that security[12][13], usability and accessibility [14] are the major challenges. Verizon 2013 data breach report showed that, 52% of the information security breaches are due to web application hacking especially cross site scripting (XSS)[15]. Acunetix in its 2016 web vulnerabilities report indicated that more than 70% of web application vulnerabilities are due to SQL injection and XSS [16].

The main objective of this study is to assess the usability, accessibility and web security vulnerabilities of Tanzania e-government websites using automatic evaluation tools in order to answer the following research questions. How high is the loading time and speed of Tanzania e-government websites in computers and laptops?, What are the accessibility standards violated by Tanzania e-government websites? and Does Tanzania e-government websites provide information security to its users?

The rest of this paper is organized as follows: Section 2 presents the country profile (Tanzania) for this study, Section 3 discusses the related work; Section 4 describes the research method used. Section 5 covers the results of usability, accessibility and web security assessment of selected e-government website. Section 6 gives the discussion of the results and finally, conclusions and future work are presented in Section 7.

## 2. COUNTRY PROFILE- TANZANIA

Tanzania, officially the United Republic of Tanzania is located in East Africa within the African great lake region and is the largest country in the region. To the north, Tanzania is bordered by Kenya and Uganda, to the west is bordered by Rwanda, Burundi and Democratic Republic of Kongo, to the south is bordered by Malawi, Zambia and Mozambique and to the east is bordered by Indian Ocean. According to Tanzania National Bureau of Statistics, the total population in 2016 is projected to be 50,142,938 [17]. Out of this population, 4,094,662 people have disabilities where difficulty in seeing is the most reported type of disability with 1,123,390 people [18]. According to Tanzania Communications Regulatory Authority (TCRA), the total number of Internet users in the country is 17.26 million [19] with an internet penetration rate of 11.5%. According to the 2016 United Nations E-Government Survey, Tanzania has a score of 0.3533 in e-government Development index (EGDI), 0.0900 in Telecomm Infrastructure Component while 0.3974 in Human Capital Component [20] rating it Medium in EGDI Level.

Despite its medium rating in E-government Development Index, the number of internet and web users continue to grow year after year in Tanzania. Some of the factors that have contributed towards this growth include an increase in the number of smart phone users, substantial decline in the cost of using internet services and the completion of the National ICT Broadband Backbone (NICTBB) [19]. Therefore, public institutions in Tanzania are required to implement a usable and accessible websites that conform to the required information security standard. So far, few studies have been conducted to evaluate the usability and accessibility of government websites in Tanzania. In [14], Content Analysis of three African countries (Kenya, Tanzania, and Uganda) was conducted based on five perspectives: website visibility, website establishment date, website ownership, website freshness, and website usability. The results show that 33 out of 37 Tanzanian government websites provided contact information; 15 provided user searching tools; 25 provided downloadable materials; while only one allowed users to submit materials online. Although this study provided useful information regarding e-government websites usability in Tanzania, no specific information was provided to facilitate improvement of e-government websites. A study to investigate whether the Websites of Tanzanian universities comply with guideline standards of accessibility and usability was conducted in [21]. The results showed that, usability and accessibility did not comply with guideline standards for the vast majority of the universities' websites. The usability of a website increases user's trust to use it. When a website is easy to use and reliable, users tend to revisit the site frequently and improves public interaction to the government [4].

A usable website must be accessible. "The power of the Web is in its universality. Access by everyone regardless of disability is an essential aspect" Tim Berners-Lee, W3C Director and inventor of the World Wide Web. Designer and developers of e-government websites must follow web accessibility guidelines in order to assure that websites are accessible to everyone. According to W3C web content accessibility guidelines, web contents is accessible if it is perceivable, operable, understandable and robust[22]. Although it is legal to design a website that is accessible by different abled users, e-government websites must provide equal opportunity to be accessible by all users by removing accessibility barriers faced by users with disabilities. Making e-government information accessible to users with disabilities is crucial for legal and ethical reasons[23].

## 3. RELATED WORK

Several studies on assessing the usability and accessibility of e-government websites in different regions in the world have been conducted [24]. The findings reported in these studies may assist web developer for e-government websites to pay attention on specific security, accessibility and usability features which are often being neglected. Website usability increases trust in e-government, but according to researchers, e-government web portals tend to have usability and accessibility problems [23]. [21] Analyzed the usability and accessibility of Tanzanian university websites and the results reported that the accessibility needs improvement to comply with W3C guidelines. A survey to investigate Web site accessibility, usability and web security of government web sites in Kyrgyz Republic was carried out in [25]. The result reported that government web sites in Kyrgyz Republic have a usability error rate of 46.3 %, accessibility error rate of 69.38 % and security vulnerabilities in these sites were revealed. Another study was launched in [26] to tests and analyze issues such as broken links, download times, time since last update, style sheets, server-side image maps, inline multimedia elements and metadata elements, browser compatibility problems, HTML validator issues and so on for several e-government web sites (Singapore, Finland, Canada, Hong Kong and Australia) for best practices. The results revealed that there are wide variations in the information and services provided by these portals as well as significant work still needed to be undertaken in order to make the portals examples of 'best practice' for e-Government services. Usability was also found to face major problems such as improper use of images, graphics and multimedia and navigation issues. [27] Used feature investigation method to investigate usability of four government websites in Uganda. The result of this study found that the investigated Ugandan websites are partially usable. Different researchers from different countries in the world have tried to assess, evaluate and analyze the usability and accessibility of e-government websites [28]–[30] for different regions.

By using WatchFire Bobby, W3C HTML validator and vsable NET LIFTA, a study was conducted to evaluate the web accessibility of e-Government websites of Saudi Arabia and Oman by adapting the W3C Web Content Accessibility Guidelines. The result reported that, accessibility of websites in both countries is low and need improvement [31]. In another study conducted in jordan using Bobby tool to evaluate the accessibility of government Web sites reported the accessibility to be very low and required an immediate improvement to reinforces user's trust. Using automated testing tools such as Websiteoptimization, Axandra and EvalAccess 2.0, [24] analyzed the usability and accessibility of 155 government websites in Malaysia and found that there are significant issues regarding usability and accessibity. Online testing tool was used to analyze the accessibility of 130 government sites in the United Kingdom according to the maturity levels defined by WCAG [32]. The results of this study indicate that only 23 % of the Web sites fulfilled the WCAG level 1 criteria and 5 % achieved WCAG 2.0 conformance. In [33] accessibility of banking websites in India was evaluated using the automatic evaluation tool based on WCAG 1.0 and WCAG 2.0 guidelines. It was found that none of the banking websites evaluated were completely accessible to people with disabilities i.e., all violated web accessibility guidelines.

Research has discovered the vulnerabilities in e-government that, more than 80% of e-government sites are vulnerable to common attacks such as XSS and SQL injection[34]. Developed countries e-governments were found more vulnerable (90%) than developing countries (50%). Another study conducted to identify e-government vulnerabilities in developing countries found that around 50% of the government websites suffer from 14 publicly know vulnerabilities including SQL injection and XSS [35]. These vulnerabilities can obstruct e-government services and lead to a significant damage to the services provided and jeopardize users' trust.

## 4. METHOD

This study assesses the usability, accessibility and web security of e-government websites in Tanzania. In particular the focus was on the key factors which affect website usability, accessibility and web security, which are loading speed, page size, loading time, broken links, accessibility errors as well as high, medium and low severity vulnerabilities. The websites assessment time was from October 1st 2016 to Dec 9th 2016. A total of seventy nine (79) e-government websites were selected from Tanzania e-government agency (http://www.ega.go.tz/) and Tanzania government website (http://www.tanzania.go.tz/) to be examined and assessed based on usability, accessibility and web security to determine their current status. Evaluating usability, accessibility and security will provide a clear insight on the current situation of e-government web portals and what should be done to improve the situation in general while maintaining information security.

The selection criteria of the websites to be examined and assessed are criteria used in [4] such as:

- The e-government websites which provide more online services to the public

- The e-government websites that are accessed by a number of users

- The websites that belong to public administration and services

To gather data for the study, automatic usability and accessibility evaluation tools were used such as google page Speed insights for loading speed analysis, Pingdom for web loading time calculation, sortsite (powemapper) for broken links and errors evaluation while web application vulnerabilities were assessed using Acunetix Web Vulnerability Scanner. To complement the results obtained using these online automatic evaluation tools, further assessment was done using WAVE tool for usability and accessibility errors, w3c link checker for broken links evaluation.

In this study, online versions of these tools were used except sortsite/powemapper and Acunetix Web Vulnerability Scanner which have the standalone applications that were installed on a desktop (window) computer. SortSite offers the ability to evaluate the Websites' accessibility, usability, compatibility with W3C standards, broken links, and search engine optimization. Acunetix web vulnerability scanner finds more security vulnerabilities than other scanner and maintain few number of false positives [16].

## 5. RESULTS

In this study, the usability was assessed by considering the loading speed, loading time and page size of the main page and number of broken links as proposed by Nielson usability guideline [36]. Web Content Accessibility Guidelines 1.0 (WCAG) such as errors present on the web pages and conformance to w3c Web Content Accessibility Guidelines (WCAG) 2.0 are used to measure the accessibility. The web application vulnerabilities measures were assessed to determine the presence of high, medium or low severity vulnerabilities mostly due SQL injection, XSS, CSRF and Denial of service (DOS). The results of the assessment are discussed in the following sub-sections.

### 5.1 Broken Links

Broken links are found in a website when some pages contain links that don't work. If clicked, such links will bring the user to a page which no longer exists. Most likely, users will land on a 404 error page, which indicates that the web server responded, but the specific page could not be found. A link may become broken for several reasons. The simplest and most common reason is that the website it links to doesn't exist anymore.

The number of broken links present in 79 websites has been evaluated using sortsite and w3c link checker and the results of those websites are categorized into four groups based on the number of broken links present as shown in Table 1.

**Table 1. Broken links**

| Number of broken links | Number of e-government websites |
|---|---|
| 0 | 0 |
| 1-5 | 43 |
| 6-10 | 22 |
| More than 10 | 14 |

The presence of broken links in a website causes navigational problem. Fig: 1 depicts each assessed websites with the corresponding number of broken links. This result indicates that 100% of 79 selected e-government websites have navigational problems due to presence of broken links.

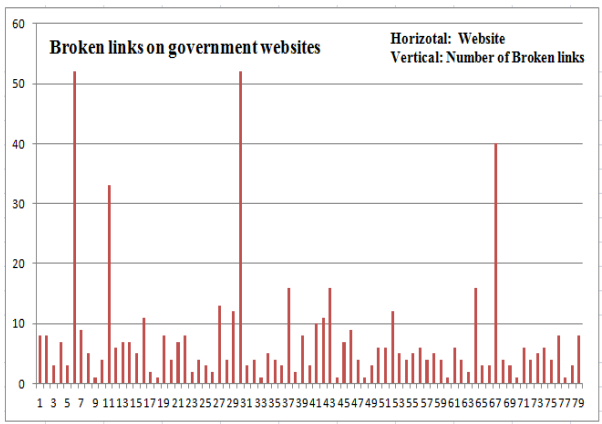

**Fig 1: Broken links on government websites**

## 5.2 Accessibility Errors

The accessibility errors are encountered by user when using a website to access the services intended for them especially by the government. Presence of errors in a website can make it difficult for user to access the services for both abled and disabled users [9]. Accessibility errors assessment was done with sortsite and wave website testing tools. These tools can identify errors present in a website. The following are errors which were found to be present in the assessed websites:

- Images without "alt" attribute
- Images with empty "alt" attribute
- Images that require a long description
- Form controls without associated label
- Reading texts on the move
- Moving or blinking content
- Links with same link text but different destinations

The total number of errors present in 79 websites is categorized into four groups depending on the number of accessibility errors present as given in Table 2.

**Table 2. Accessibility Errors**

| Number of Errors | Number of e-government websites |
|---|---|
| 0 | 0 |
| 1-20 | 13 |
| 21-50 | 43 |
| more than 50 | 23 |

Some errors appeared to be due to inconsistency in the language of the interface. After identifying the pages with spelling errors, the author tried to open these pages manually and found that, link at the English interface opened the page that displayed Swahili and English contents or Swahili contents only. There are two main accessibility outputs for SortSite tool, better than average or worse than average, based on global standards. In this study, the majority of websites' accessibility is worse than average (62%). Fig: 2 show the distributions of selected e-government websites and the number of accessibility errors present.

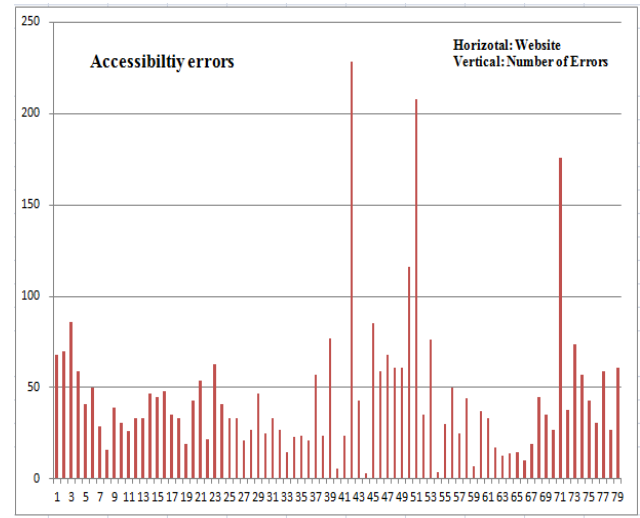

**Fig 2: Accessibility errors on government websites**

## 5.3 Website Loading Speed

The loading speed of 79 selected websites has been assessed and examined using google speed insight online tool. The result of the loading speed assessment is divided into three groups (rating) as shown in Fig: 3. The page loading speed score ranges from 0 to 100%. A higher score is better and a score of 90% or above indicates that the page is performing well [37]. The result indicates that, the vast majority of websites (56 out of 79) have a low loading speed for their main page.

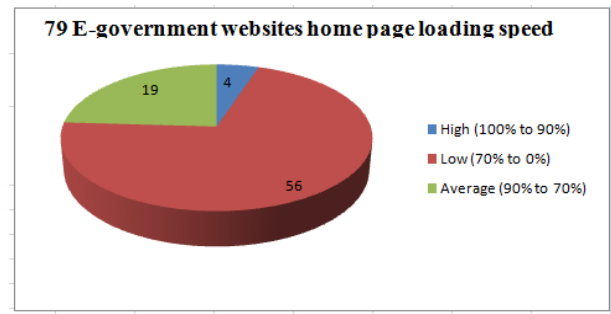

**Fig 3: E-government websites main page loading speed.**

## 5.4 Website Loading Time

The loading time of 79 selected website has been assessed by using pingdong online tool. Loading time is the time taken to download to the browser the web pages of selected websites. Very slow websites (of 10 seconds and more) negatively impact the user of the website. The loading time is major contribution factor to page abandonment by the user as they lose patience for a page that takes too long to load [37]. Slow web pages response time results in an increased page abandonment. Computer system response time to a user's request or inquiry at a terminal was proposed to be 2 seconds back in 1960's [38]. In the today's information society, computer and internet speed is improved as compared to the past years. In [39] a study conducted came up with findings that indicate the amount of time a user can wait for a web page to load. Therefore, this study uses the results obtained in [39] to set up the cluster for comparing different web loading time. Rating for loading time is divided into three rating as shown in Table 3.

**Table 3. Loading Time**

| Rating | Loading Time | Percentage of websites (%) |
|---|---|---|
| High (Good) | <=5 seconds | 34.18 |
| Average | <=10seconds | 27.85 |
| Low (poor) | >10 seconds | 37.97 |

According to [4], high loading speed lead to faster response time to the user and simplify browsing. The authors here also stressed that, the average loading speed makes the user wait a bit but still the user satisfying with speed. More than 37% (52 out of 79) of the Tanzania e-government websites have uploaded time duration of more than 10 seconds as shown in Fig. 4. The important factors that mainly contribute to the poor loading speed of websites are server response time, page size, large embedded multimedia contents, incompatible images etc.

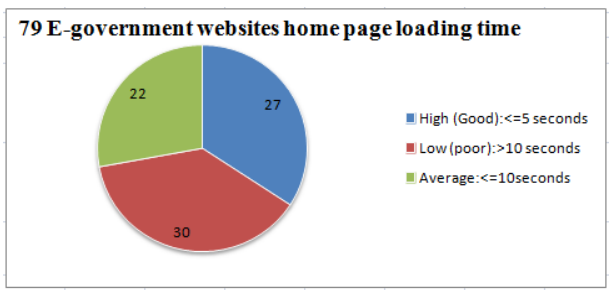

**Fig 4: E-government main page loading time**

## 5.5 Web page size

The web page size of selected 79 websites shown in Table 4 has been assessed by using pingdong online tool. An average small web page is approximately 12 KB [40], and that will load very quickly. The more media on a page, the bigger the page size cause a website to load slowly. Reasons for slower website may be due to embedded videos, images, audio, graphics, flash, and other forms of media that increase the page size of a website. A website with a page size of up to 2MB is said to have an average size [41].

**Table 4. Page Size**

| Ratings | Page size | Number of e-government websites |
|---|---|---|
| High (Fast) | <=12kb | 0 |
| Average (good) | <=2MB | 44 |
| Low (slow) | More than 2MB | 35 |

A fast website give a good user experience and satisfies users while a very slow websites is a bad user experience. Fig: 5 shows the percentage distribution of the page sizes for the main page of 79 selected e-government websites. How fast a website loads is critical in maintaining user. The results indicate that, 35 out of 79 of the Tanzania e-government websites have main page size, uploaded in more than 12 KB. These results prove that usability issues are less than being considerate in Tanzania e-government web design by website developer.

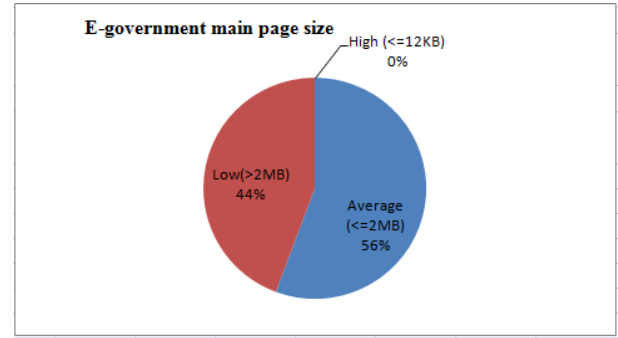

**Fig 5: E-government web sites main page size**

## 5.6 Web Security Vulnerabilities

Number The security vulnerabilities present in 79 selected website has been assessed by using Acunetix Web Vulnerability Scanner tool and the result is shown in Table 5.

**Table 5. Website Vulnerabilities**

| Type of vulnerability | Number of e-government websites | Percentage of vulnerable websites (%) |
|---|---|---|
| XSS only (High) | 23 | 29.11 |
| SQL injection only (High) | 4 | 5.06 |
| XSS and SQL injection (High) | 13 | 16.46 |
| CSRF and DOS (Medium) | 51 | 64.56 |

As it can be seen in Table 5, 40 out of 79 (50.6%) assessed websites have one or more high-severity vulnerability (SQL injection-sqli, cross site scripting-XSS) while 51 0ut of 79 (64.5%) have one or more medium-severity vulnerabilities (Cross site request forgery-CSRF, Denial of Service-DOS). Therefore, some sites appeared to have both high-severity and medium-severity vulnerabilities as shown in Fig: 6. High-severity vulnerability gives an attacker ability to compromise the confidentiality, integrity or availability of a website without specialized access, user interaction or circumstances that are beyond the attacker's control [16]. This makes it very possible for an attacker to gain access to other systems on the internal network of the vulnerable website. Medium severity vulnerabilities cause an attacker to have a partial ability to compromise the confidentiality, integrity or availability of a target system. Attacker can use this weakness together with high-severity vulnerability to simplify the attack.

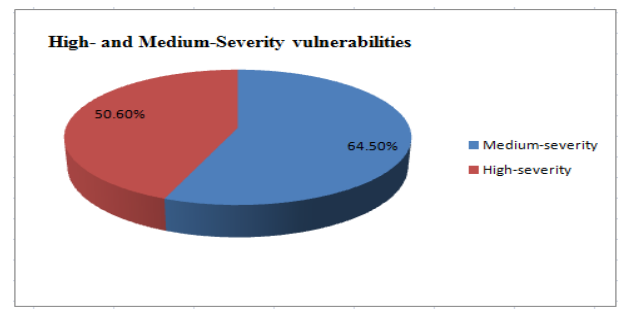

**Fig 6: Vulnerabilities in the e-government web sites**

Tanzania public institutions need to be clear on what this means. High-severity vulnerability could allow attackers to gain unauthorized access to data and systems, potentially to sensitive financial, customer, health data and trade secrets. They could also move to other systems to escalate the attack even further. Web Application (High-severity) lowers user's trust and willingness of using e-government websites.

## 6. DISCUSSION

Usability, accessibility and web security assurance to Tanzania e-government websites should be the key features for the services provided to the citizens. In this study, the selected government websites have been assessed using online evaluation tools. Assessment were conducted based on usability, accessibility and web security of a particular website. Each type of assessment was conducted using a specific online diagnostic tools and/or desktop tools. Assessment included broken links, accessibility errors, loading speed and time for the main page, page size of the main page and severity of web vulnerabilities ( high, low and medium).

In general, considering the selected government websites, it can be said that usability is given a very low priority in Tanzania e-government websites, with 100% of site having broken links, 35 out of 79 of the Tanzania e-government websites' main page size is more than 12 KB, 56 out of 79 have a low loading speed for their main page and more than 37% of the Tanzania e-government websites have uploading time duration of more than 10 seconds.

The accessibility assessment showed that 100% of government websites have accessibility errors and do not conform to w3c Web Content Accessibility Guidelines (WCAG) 1.0/2.0. None of government websites found to have conformance level A of W3C. Nevertheless, 100% of government websites failed to pass priority 1 checkpoints for accessibility errors as 100% have errors. These results suggest that w3c Guidelines WCAG 1.0 and WCAG 2.0 should be considered.

Thus, during development, improvement has to be made to make e-government websites usable and accessible and convenient to use by their users. Tanzania e-government agency (ega) should come up with a frameworks that every public institutions must follow in developing their websites.

The main security concern about government web sites is that 50.6% have one or more high-severity vulnerability (SQL injection-sqli, cross site scripting-XSS) while 64.5% have one or more medium-severity vulnerabilities (Cross site request forgery-CSRF, Denial of Service-DOS). Some sites appeared to have both high-severity and medium-severity vulnerabilities and have passwords transmitted over HTTP. Government websites are often the target for hackers, therefore these vulnerabilities are real threats to the security and privacy of information shared by users and the government. If an attacker takes advantage against these weaknesses, it could simplify the attack. Therefore, web developers should make sure that passwords are encrypted over HTTP and there is proper input validation to safeguard websites against SQL injection and XSS.

## 7. CONCLUSION AND FUTURE WORK

The results of this work show that there is a need to improve the usability, accessibility and security of Tanzania e-government websites as it saves all kind of citizens (both with and without disabilities). Although some of the problems identified such as broken links, loading speed and time of the home page could be fixed after deployment, problems such as accessibility errors and web vulnerabilities must be fixed during design and development stages. Fixing these problems during development cost less and is simpler as compared to fixing it after deployment. Accessibility errors identified by sortsite tool can be solved by following design procedures that could eliminate these errors and conform to w3c Web Content Accessibility Guidelines (WCAG) 1.0. Errors such as images without "alt" attribute, images with empty "alt" attribute, reading texts on the move, moving or blinking content and links with same link text but different destinations should be removed during development stages.

Since the results revealed that web developers do not consider accessibility and usability during development stages, there is a need for training and awareness for web developers and designers on how to develop websites that comply to usability and accessibility standards. All web sites that comply to these standard are easier to develop, maintain and update.

As the number of smart phone users increase in Tanzania, the future works is to evaluate and analyze the usability and accessibility of Tanzania e-government web sites in mobile devices. Also from these results, user experience on accessibility of Tanzania e-government websites need to be evaluated.

## 8. ACKNOWLEDGMENT
The author acknowledge the University of Dodoma, College of Informatics and Virtual Education for all the supports he received to allow this research to be undertaken.